\documentclass[aps,onecolumn,showpacs,preprintnumbers,amsmath,amssymb]{revtex4}
\usepackage{graphicx}

\begin{document}
\draft
\title{Renewal and memory properties in the  random growth of surfaces}
\
\author{R.Cakir}

\author{P. Grigolini}
\email{grigo@unt.edu}

\author{M. Ignaccolo}

\affiliation{Center for Nonlinear Science, University of North Texas, P.O. Box 311427, Denton, Texas 76203-1427, USA }

\date{\today}

\begin{abstract}

We use the model of ballistic deposition as a simple way to establish cooperation among the columns of a growing surface, \emph{the single individual of the same society}. We show that cooperation generates memory properties and at same time non-Poisson renewal events. The variable generating memory can be 
regarded as the velocity of a particle driven by a bath with the same time scale, and the variable generating renewal processes is the corresponding diffusional coordinate.

\end{abstract}
\pacs{05.40.-a, 05.65.+b, 89.75.Da}
\maketitle

\section{Introduction}\label{introduction}

The Science of Complexity \cite{complexity1} is an emerging field of research, which is attracting an increasing number of investigators. There is a general agreement that 
the deviation from the exponential behavior is a sign of complexity. However, on the origin of this deviation is there are different proposals. There are authors emphasizing the memory properties \cite{lee,usatenko,min} and other \cite{sokolov1,sokolov2,subordination1,subordination2,subordination3,budini} stressing the non-Poisson renewal properties of complex systems. 

As an attractive example of non-Poisson renewal processes we refer to the field of single-molecule spectroscopy \cite{singlemolecule} and blinking quantum dots \cite{bqd1,bqd2,bqd3} . There is an increasing evidence that phenomena such as intermittent fluorescence yield histograms of the "on" and "off" sojourn time distribution that depart dramatically from the exponential condition. At the same time, a careful statistical analysis proves that these are renewal processes \cite{bqd3,paradiso}. Renewal theory \cite{renewal} describes, for example, successive replacements of light bulbs: when a bulb fails it is immediately replaced by a new one, which works with no memory whatsoever of the time duration of the earlier light bulb. Thus, renewal processes are characterized by the occurrence of events that reset to zero the system's memory. In which sense, therefore, a non-Poisson renewal process is a manifestation of organized behavior, which seems to imply \cite{lee,usatenko,min} long-time memory? 

It is tempting to express the non-exponential properties of a complex system in terms of ordinary Poisson processes with a fluctuating Poisson parameter \cite{superstatistics1,superstatistics2,superstatistics3}.
Using this perspective, it is possible to prove \cite{abe} that complex networks emerge from fluctuating random graphs. However, it is worth remarking that slow fluctuations \cite{barbi} reduce the occurrence of renewal events and generate memory properties compatible with the adoption of stationary correlation functions.

Another example of non-Poisson renewal process is given by the fields of seismic fluctuations \cite{mega}, where, however, there is no general agreement on the renewal character of the process.  
 The authors of Ref. \cite{mega} argue that
the distribution of time distances between two consecutive main shocks corresponds to the properties of a non-Poisson renewal process. Is this conclusion reliable, or does it conflict with the notion of main shocks being complex processes,  are \cite{predictable}, to some extent, predicable?
Another example, of greater interest for this paper, is given by the random growth of surfaces. The authors of a recent work \cite{arne} have used the concepts of non-Poisson renewal theory to describe and derive the same properties that other authors \cite{others1,others2,others3}
derive from the adoption of the Fractional Brownian Motion (FBM) perspective \cite{fbm}. On one side, we have renewal, with jumps resetting to zero the system's memory, and on the other we have fluctuations with infinite memory. Which is the best signature of complexity?

The purpose of this paper is to prove that both aspects are present in the processes of random growth of interfaces, which are a natural realization of of the Renewal Cooperation (RC) perspective proposed in the earlier work of Ref. \cite{cakir1}. 

The outline of the paper is as follows. In Section \ref{ballisticmodel} we shall give detail on the model used for the illustration of our approach to complexity, and we shall define the variable $\tilde \xi(t)$ which is the signature of memory, and the variable $y$, whose origin regression is a non-Poisson renewal  process. 
In Section \ref{aging} we establish the renewal character of the variable $y$ and in Section \ref{memory} we study the memory properties of the variable $\tilde \xi(t)$. In Section \ref{final} we shall point out the result of this paper. 

\section{Ballistic Model} \label{ballisticmodel}

\begin{figure}[!b]
\begin{center}
\label{f1}
\hspace {-1cm}
\includegraphics[angle = 0,scale = 0.4]{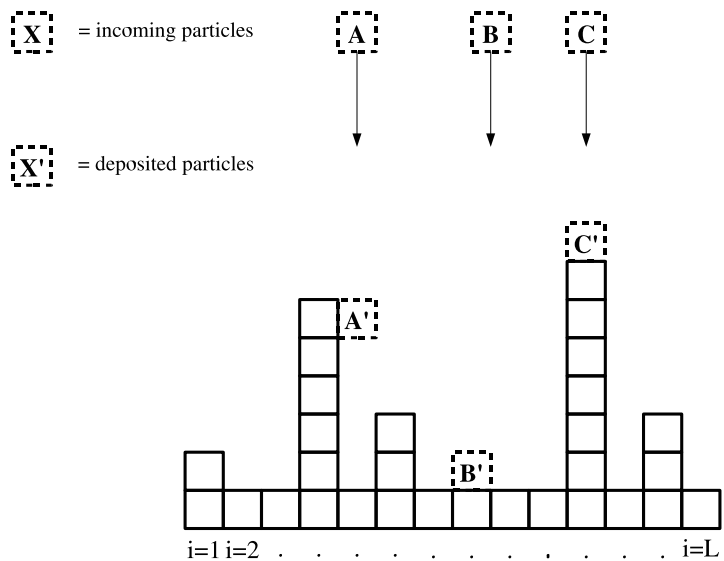}
\caption{Model of ballistic deposition. The particle $B$ settles at 
the top of an earlier particle of the ame coloum, given the fact that 
there are no particles at a higher level in the two nearest neighbor 
columns. The particle $A$ sticks to the right side of the left 
nearest neighbor colomn rather than at the top of a particle in the 
same colomn at a lower level.}
\end{center}
\end{figure}

We plan to shed light into the issues mentioned in Section \ref{introduction} with the model of Ballistic Deposition (BD), which, is, in fact, a paradigmatic model for complexity.  We shall limit our discussion to the (1+1)-dimension case, and as a consequence, Fig.1 is fully adequate to  illustrate how the model works. At any time step, $n=1, 2,  \cdot \cdot \cdot$  we select randomly one of the $L$ columns, and we drop a particle on it. The particles fall down in sequence till to settle either at the bottom of the column or at the top of an earlier selected particle that by chance fell down in the same column. However, if the height of one of the two nearest neighbor columns is higher than the selected column, the particle sticks to the side of the highest particle of this neighbor column. There is also periodic boundary conditions to decrease the effect of limited size of surface.

Actually, this side sticking action corresponds to a transverse transport of information, through which the column under study is informed about the height of the surrounding columns. Thus, examining the time evolution of a single column is equivalent to studying the behavior of a single individual and to assessing to what an extent it reflects the properties of the whole society.  We plan to prove that the cooperation among the different columns generates memory and this memory generates renewal effects. 

It is well known \cite{barabasi} that the time distance between the arrival of one particle in this column and the next has the Poisson time distribution
\begin{equation}
\label{poisson1}
\psi(\tau) = \lambda \exp(-\lambda  \tau),
\end{equation}
where 
\begin{equation}
\lambda = \frac{1}{L}.
\end{equation}
Of course, when one particle arrives, the height of the column increase by a quantity that can be also much greater than $1$, thank to the side-sticking effect. We denote by $\xi(t)$ the height increase of the column, and this quantity is $0$ when no particle arrives, and a number equal to $1$ or larger when a particle arrives.

The height of the column at a give time $t$ is given by
\begin{equation}
h(t) = \sum_{n = 0}^{t} \xi_{n},
\end{equation}
with $\xi_{0} = 0$. 
To produce a renewal process we use the variable 
$y(t)$ defined by
\begin{equation}
y(t) \equiv h(t) - <h(t)>,
\end{equation}
where $<h(t)>$ denotes the average over all the sample columns. In the earlier work of Ref. \cite{arne} it was argued that the regressions of the variable $y$ to the origin $y = 0$ is a renewal process. We shall prove this property with a compelling numerical experiment.

Let us now define the variable that is the signature of the memory properties generated by the BD model. This is the variable $\tilde\xi(t)$ defined by
\begin{equation}
\label{rasitbecareful}
\tilde \xi \equiv \xi(t) - \bar \xi(t),
\end{equation}
where the symbol $\bar \xi$ denotes a time average. We generate a very long, but finite, sequence $\xi(t)$, and we define the time average
\begin{equation}
\bar \xi = \frac{\int_{0}^{T_{max}} dt \xi(t)}{T_{max}}.
\end{equation}

\begin{figure}[!ht]
\begin{center}
\label{f2}
\hspace {-1cm}
\includegraphics[angle = 270,scale = 0.5]{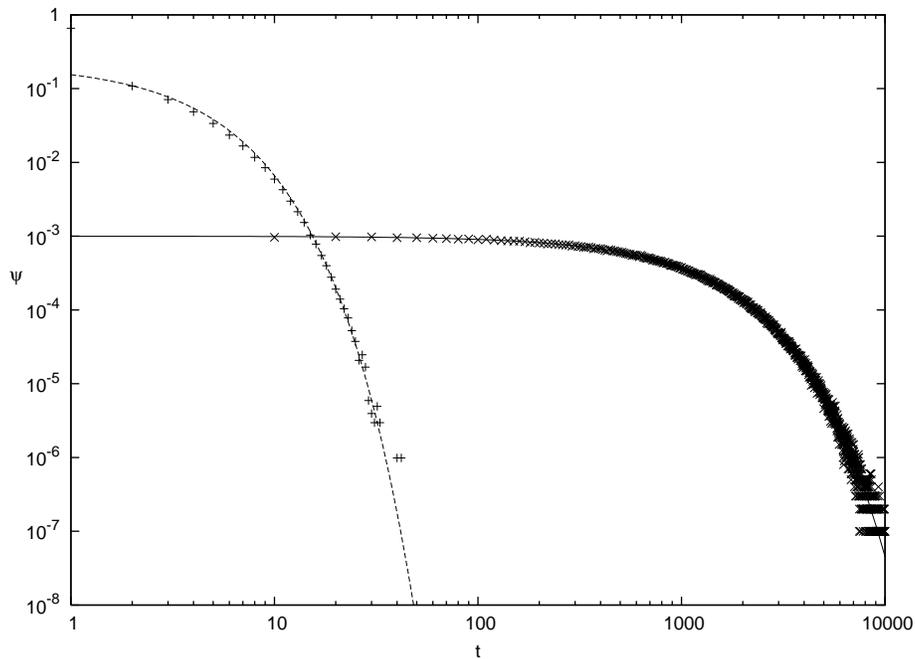}
\caption{Probability distribution of jumps, $\xi$, (+), fitting to 
dashed line  $0.22~exp(-0.35~t)$; and waiting times of jumps, (x), 
fitting to solid line  $0.001~exp(-0.001~t)$. L = 1000}
\end{center}
\end{figure}

Fig. 2 illustrates the distribution density  $\psi(\tau)$ and 
$\psi(\xi)$, and prove that the latter distribution, as well as the former, is an exponential,
\begin{equation}
\label{poisson2}
\psi(\xi) = a \exp(-b \xi).
\end{equation}

On the basis of these results one would be tempted to conclude that the single column process is Poisson. We shall prove that it is not so because the fluctuation $\xi(t)$ has memory, in spite of its exponential distribution. This memory is a consequence of the process of transverse transport of information, and it can be considered a signature of cooperation. 

\section{aging} \label{aging}
The authors of Ref. \cite{arne}  studied the variable $y$, whose variance $w(L,t)$, in accordance with the literature, was shown to obey 
the anomalous prescription
\begin{equation}
\label{varianceofy}
w(t) \propto t^{\beta},
\end{equation}
with $\beta < 0.5$. Actually, for $L \rightarrow \infty$ the power index $\beta$ is expected to fit the Kardar-Parisi-Zhang (KPZ) prediction \cite{kpz} $\beta = 1/3$. To speed up the numerical calculation we limit our calculations to the case of $L = 1000$, and consequently to the value $\beta =0.28 $.  The authors of Ref. \cite{arne} argued that the $y$-process can be reproduced with remarkable accuracy by adopting the subordination approach \cite{sokolov1,sokolov2,subordination1,subordination2,budini}. It is important to stress that the variance $w(t)$ is a Gibbs property, obtained by making a statistical average on all the columns of the sample. The single trajectory $y(t)$ is derived from the stochastic trajectory
of $y(n)$, which is driven in the natural time scale by the ordinary Langevin equation
\begin{equation}
\frac{d}{dn} y = - \gamma y(n) + f(n),
\end{equation}
and assuming that the transition from the natural time scale $n$ to the $t$-time scale is realized with the prescription $t(n+1) - t(n) = \tau(n)$, where $\tau(n)$ is randomly derived from a distribution density with power law $\mu < 2$. This is an approach well distinct from the ordinary memory approach \cite{lee,usatenko,min}, which is based on a generalized Langevin equation, whose memory kernel is a conventional equilibrium correlation function. It is also important to point out that
there exists a connection  between $\mu$ and $\beta$ 
stemming from It is worth remarking that all this is also a clear evidence of the fact that the relationship
\begin{equation}
\label{ling}
\beta = 2 - \mu,
\end{equation}
which was derived by the authors of Ref. \cite{arne} from the renewal theory, rather than by using the FBM theory, as done by earlier investigators \cite{lingandding}. 

\begin{figure}[!b]
\begin{center}
\label{f3}
\hspace {-1cm}
\includegraphics[angle = 270,scale = 0.5]{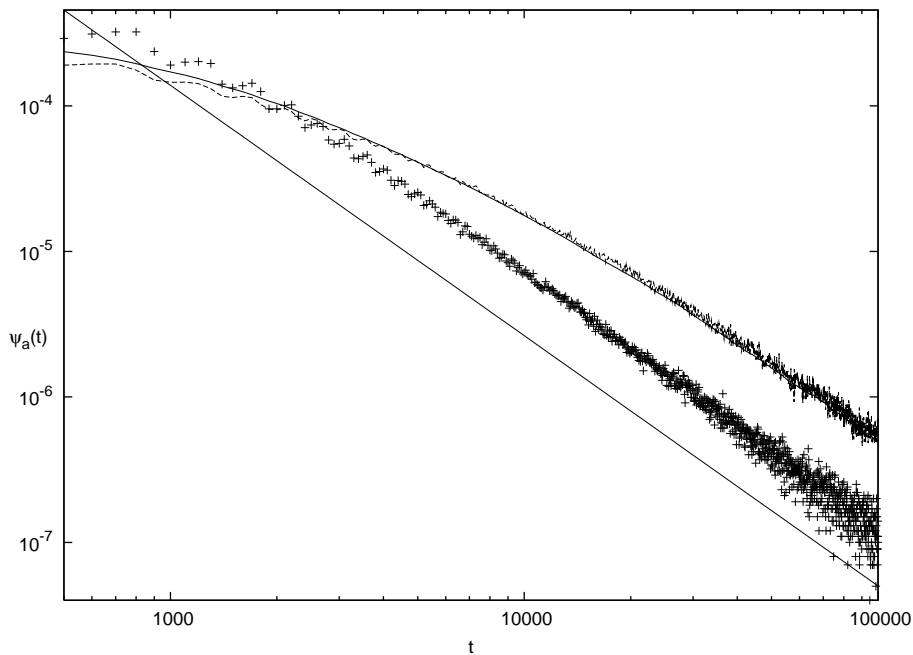}
\caption{Probability distribution of waiting times of $y(t)$ are 
shown with (+) when there is no aging. The straight solid line is 
$20~x^{-1.72}$. Probability distribution of aged waiting times are 
dashed curve for experimental and solid curve for theoretical 
calculation.}
\end{center}
\end{figure}

The authors of Ref. \cite{arne} established a very good agreement between theory and numerical simulation by adopting this subordination perspective. 
In this section we confirm the arguments of Ref.\cite{arne} by means of an aging experiment, recently proposed to prove the renewal character of BQD processes \cite{paradiso}. According to this procedure, we study first the recursions to the origin of the fluctuation $y(t)$, which are described by Fig. 3. We see that $y(t)$ remains in the positive portion of the $y$-axis for a given amount of time, then $y$ moves to the negative portion of this axis, remains there for a while, till to re-cross the origin $y=0$, and so on. We evaluate the histogram and we find an inverse power law with $\mu = 1.72 $. According to the prescription of Ref. \cite{paradiso} we evaluate also the aged waiting time distribution $\psi^{(exp)}(\tau, t_{a})$. This is done as follows. The re-crossing times generates the time series
$\{t_{i}\}$. For each time $t_{i}$ we record the first time
of the sequence at a distance from $t_{i}$ equal to or larger than $t_{i} + t_{a}$. This time will be $t_{k}$, with $k > i$. We record the time distance $\tau(t_{i}, t_a) = t_{k} -(t_{i} + t_{a})$. We repeat the procedure for all the times of the sequence $\{t_{i}\}$, and use the sequence of these recorded time distances to generate the distribution density $\psi^{(exp)}(\tau, t_{a})$. We have also to evaluate 
\begin{equation}
\psi^{(theor)}(\tau, t_{a}) = \frac{\int_{0}^{t} dy \psi(t+y)}{K(t_{a})},
\end{equation}
where $K(t_{a})$ is a suitable normalization constant and $\psi(t)$ is the experimental waiting time distribution corresponding to $t_{a} = 0$. 
The accordance between $\psi^{(theor)}(\tau, t_{a})$ and $\psi^{(exp)}(\tau, t_{a})$ is judged \cite{paradiso} to be the numerical evidence of the renewal nature of the process under study. 

Fig. 3 confirms the renewal nature of the $y$-process, thereby providing further support for the theoretical approach to the stochastic growth of a single column proposed in Ref. \cite{arne} and for the adoption of the renewal theory to derive the crucial relation of Eq. (\ref{ling}). Using Eq. (\ref{ling}) and the numerical  results of Fig. 3 for $\mu$, we obtain $\beta = 0.28$. This same power index will be derived from the memory properties discussed in Section \ref{memory}. 

\section{memory} \label{memory}
In this section we adopt a single-trajectory perspective, rather than the Gibbs approach of Section \ref{aging}. 
How can we reveal the cooperative nature of the process if we limit ourselves to studying the time evolution of only one column? Apparently, the statistical analysis that led us to Eqs. (\ref{poisson1}) and (\ref{poisson2}) would suggest that the anomalous growth property of the interface as a whole is annihilated by the observation of a single column. 

To assess the memory properties created by cooperation, we proceed as follows. We create the diffusing variable
\begin{equation}
\label{diffusingvariable}
x(t) = \int_{0}^{t} \tilde \xi(t^{\prime}).
\end{equation}
Then,  using the proposal made by Stanley and co-workers years ago \cite{stanley1,stanley2} we convert the single diffusing trajectory of Eq. (\ref{diffusingvariable}) into many diffusing trajectories.  This is done as follows.  We consider a window of size $l$, and we move it along the sequence. We record the space positions of the random walker at times $s$  and $s+l$. This makes it possible for us to define the $s$-th random walker that in the time $l$, moves from the origin $x=0$ to the position $x(s,l) = x(s+l)- x(l)$. 
Thus we have a set of walkers, each  walker denoted by the label $s$, moving for a time $l$ and covering the distance $x(s,l)$. We imagine all these walkers to depart from the origin $x= 0$ at the same time. This means that the set of values   $x(s,l)$ determines a distribution, whose variance $w(l)$ is given by
\begin{equation}
\label{variance}
w(l) = \left < \left (x(s,l) - <x(s,l)> \right )^{2} \right >^{1/2},
\end{equation}
the averages $<...>$ being made on all the labels $s$. It is evident that if the single column reflects the complex behavior of the whole surface, see Eq. (\ref{varianceofy}),
then
\begin{equation}
\label{asymptoticvariance}
\lim_{l \rightarrow \infty} \log( w(l)) = \beta \log (l).
\end{equation}
Fig. 4 shows that this prediction is fulfilled and that, as expected, $\beta = 0.28$. 

\begin{figure}[!ht]
\begin{center}
\label{f4}
\hspace {-1cm}
\includegraphics[angle = 270,scale = 0.5]{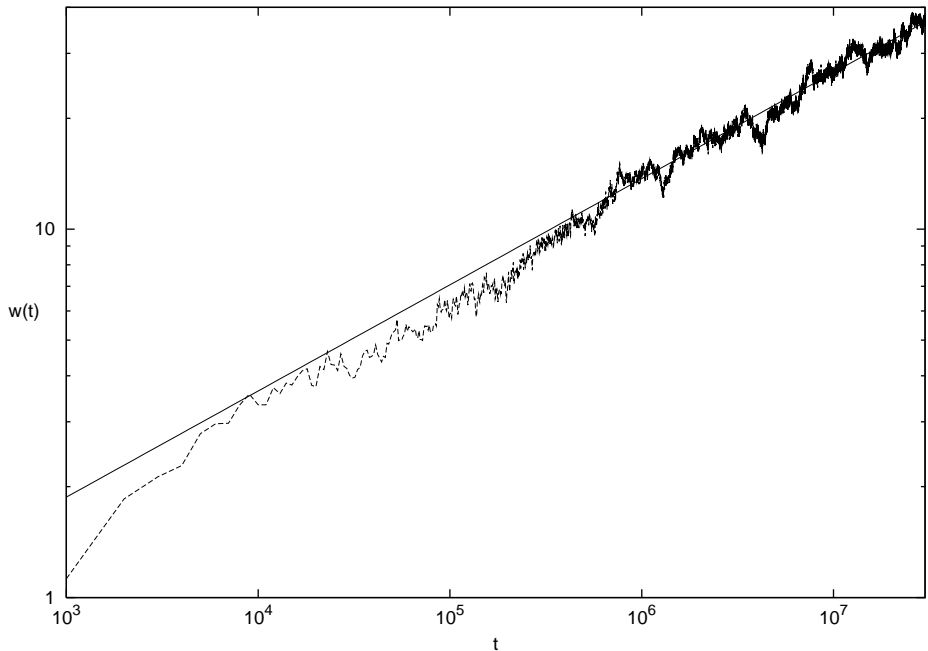}
\caption{Variance of $x$ is shown with dashed line. The solid line is 
$0.256~x^{0.288}$}
\end{center}
\end{figure}

Under the stationary assumption, it is straightforward to prove (see, for instance, Ref. \cite{labella}) that
\begin{equation}
\label{secondmoment}
\frac{d}{dt}<x^{2}(t)> =  <\xi^{2}> \int_{0}^{t} \Phi_{\tilde \xi}(t') dt',
\end{equation}
where $\Phi_{\tilde \xi}(t')$ denotes the equilibrium correlation function of the fluctuation $\xi(t)$. 
In the time asymptotic limit, which is the intermediate regime illustrated by both Fig. 3 and Fig. 4, with no upper bound time limit, determined by the third regime, we get
\begin{equation}
\label{scaling}
w(t)^{2} = <x^{2}(t)> \propto t^{2\beta},
\end{equation}
namely, we reiterate that the diffusion process has anomalous scaling defined by $\beta$, with $\beta < 1/2$. 
By differentiating Eq. (\ref{secondmoment}) with respect to time we establish a connection with the time asymptotic properties of  $\Phi_{\tilde \xi}(t)$. Due to the fact that $\beta < 1/2$ we conclude that the correlation function $\Phi_{\tilde \xi}(t)$ must have a negative tail, namely that 
\begin{equation}
\label{negativetail}
\lim_{t \rightarrow \infty}  \Phi_{\tilde \xi}(t)  = - \frac{constant }{t^{\psi}},
\end{equation}
with
\begin{equation}
\label{power}
\psi = 2 - 2 \beta.
\end{equation}

This property establishes a closer connection with the earlier work of Ref. \cite{cakir1}, where the variable responsible for memory yields a correlation function with a negative tail. In the case of Ref. \cite{cakir1} the correlation function is known theoretically, so that it is possible to move from the correlation function to the variance time evolution using Eq. (\ref{secondmoment}). In the present case, we do not have available any analytical approach to the equilibrium correlation function of $\tilde \xi(t)$.  Its numerical derivation, as pointed out by Eq. (\ref{secondmoment}), would be equivalent to differentiating twice the variance $w(t)$, which is numerically a source of big errors. In, fact, the numerical approach to the equilibrium correlation function, not shown here, yields a negative tail, but the assessment of the correct power requires rich statistics and excessive computational time.

Thus, to do double check our conclusion, that the variable $\tilde \xi(t)$ is the source of memory, with no renewal events, we prefer to accurately examine the physically plausible conjecture made by the authors of Ref. \cite{spagnolo} that the non-Poisson renewal events may be hidden among the pseudo-Poissonian fluctuations of the variable $\xi(t)$. According to this conjecture, either the fluctuations of $\xi(t)$ above a given threshold $R$ or the time distances between two consecutive fluctuations of $\xi(t)$, exceeding the threshold $R$, may deviate from the exponential distribution. Fig. 5 and Fig. 6 illustrate these two conditions, and prove that in both cases the asymptotic time behavior is exponential. 

We therefore conclude that there is no evidence of renewal events in the dynamics of $\tilde \xi(t)$. 
The stationary assumption made to deal with the correlation function of $\tilde \xi(t)$ does not conflict with the non-stationary nature of $y$. This is so because $\tilde \xi(t)$ and $y(t)$ are the signatures of cooperation and renewal, respectively. Cooperation generates a slow decaying correlation function, but the diffusion process generated by these fluctuations is renewal, in accordance with Ref. \cite{cakir1}. 

\begin{figure}[!ht]
\begin{center}
\label{f4}
\hspace {-1cm}
\includegraphics[angle = 270,scale = 0.5]{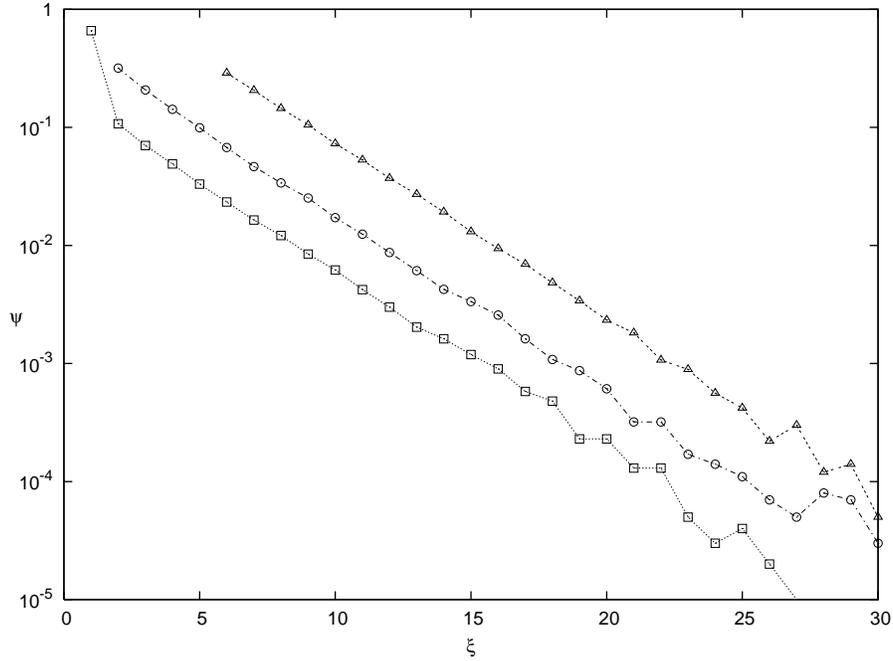}
\caption{Probability distribution of $\xi>$R, for R=0 (square), R=1 (circle) and R=5 (triangle).}
\end{center}
\end{figure}
\begin{figure}[!ht]
\begin{center}
\label{f4}
\hspace {-1cm}
\includegraphics[angle = 270,scale = 0.5]{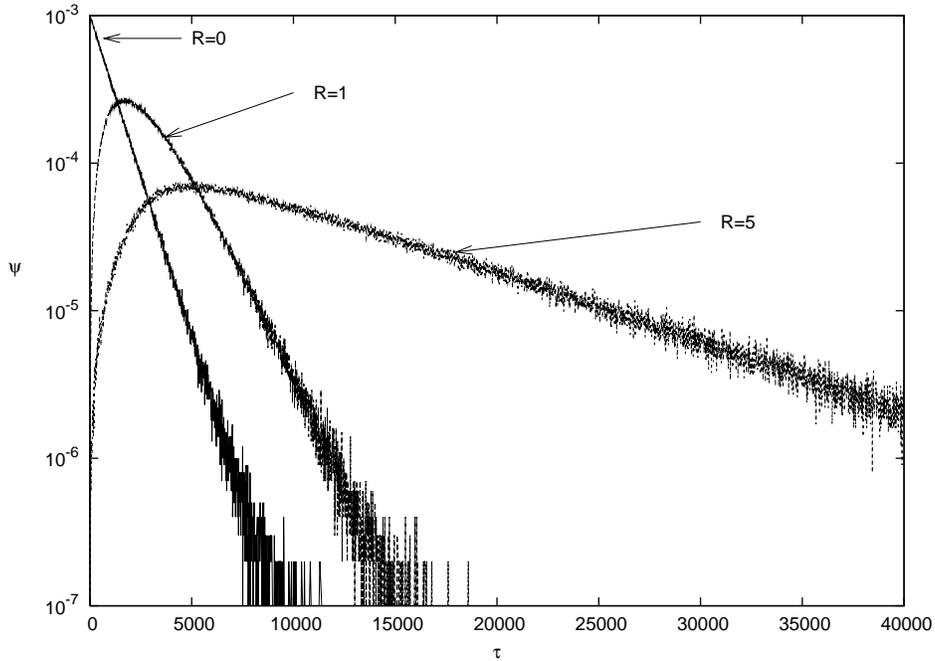}
\caption{Probability distribution of waiting times of $\xi>$R, for R=0, R=1 and R=10.}
\end{center}
\end{figure}

\section{concluding remarks}\label{final}
This paper shows how to reconcile a perspective based on memory with one based on renewal non-Poisson processes. To a first sight, these two visions can be perceived as being incompatible. For instance, the adoption of the FBM theory adopted by some authors to account for persistency \cite{persistency1,persistency2,persistency3} seems to conflict with the renewal approach. On the contrary, in accordance with the recent result of Ref. \cite{cakir1} this paper shows that the cooperative nature of the BD model generate memory and, consequently, anomalous diffusion. However, the origin re-crossings of the variable $y$ are renewal, in the same way as the diffusion variable generated by the long-memory fluctuations used to generate dynamically FBM exhibit a renewal character \cite{cakir1}.

We acknowledge financial support from Welch foundation through Grant N. 70525 and from ARO through Grant No. W911NF-05-1-0059. 

\end{document}